\documentclass[journal]{IEEEtran}
\usepackage{amsmath}

\input{tcilatex}

\begin{document}

\title{Bayes-optimal detection of TNT content by nuclear quadrupole resonance%
}
\author{Sven Aerts$^{1,\ast }$, Dirk Aerts$^{1,\dagger }$, Franklin Schroeck$%
^{2,\ddagger }$, and J\"{u}rgen Sachs$^{3,\sharp }$ }
\maketitle

\begin{abstract}
We study the statistical performance and applicability of a simple quantum
state discrimination technique for the analysis of data from nuclear
quadrupole resonance experiments on a TNT sample. The target application is
remote detection of anti-personnel landmines. We show that, even for data
that allows the determination of only one time dependent component of the
NQR\ subsystem, the use of the Bayes optimal detector leads to greatly
improved ROC curves with respect to the popular demodulation technique,
especially for spin echo signals with a low signal to noise ratio. The
method can easily be extended to incorporate results from other sensing
modalities and the incorporation of informationally complete measurements
that estimate the full density matrix of the NQR subsystem.
\end{abstract}

\section{Introduction}

In this report we address the problem of deciding whether a given closed
volume contains TNT content by means of nuclear quadrupole resonance
measurements. Nuclear quadrupole resonance (NQR) signals result from the
relaxation of nuclear quadrupole momenta to their original thermal
equilibrium position after an initial, high power RF pulse has been applied.
The thermal equilibrium configuration of the nuclear spins is a function of
the electromagnetic field in the vicinity of the quadrupole active nuclei.
As a result, the NQR spectrum is very specific with respect to chemical
compounds in the substance involved and can serve as a fingerprint to
identify that substance. Because of its high potential value in remote
explosive detection, there is renewed interest in NQR\ methods for landmine
and UXO detection, as well as for securing high risk areas such as airports
by non-intrusive means. Quite incidentally, NMR and NQR systems have
recently also received great attention for their applicability in the fast
growing field of quantum information \cite{Nielsen huang} and state of the
art quantum computers are currently based on NMR. In NMR, a high intensity,
homogenous magnetic field introduces a preferred axis of quantization and
causes the energy levels to undergo a Zeeman splitting. In NQR, this
splitting is caused by the interaction of the nuclear quadrupole with the
electric field gradient. One would then conjecture that a quantum
statistical analysis of the data is optimal \cite{Helstrom}, \cite{Holevo}
and \cite{Malley and Hornstein}. However, the macroscopic bulk size and
consequently large number of spins necessary for an appreciable signal
strength, as well as the far from absolute zero temperature of the sample in
realistic conditions, call for a classical statistical approach\footnote{%
A\ similar discussion about the physical nature of NMR and NQR systems is
held in the quantum information community. Whereas it is well-known that the
magnetization of single spin systems can be recovered by a classical
spinning top, it seems quantum principles are called for to explain the
coherence effects we encounter when a large number of spin systems interact.
In absense of an experimental demonstration of entanglement in NMR systems,
it has been argued that current NMR based quantum computers are not, in
fact, real quantum computers. The opponents of this argument maintain that
genuine quantum information processing tasks have been performed on NMR
systems, albeit not very efficiently, and that perhaps entanglement is not
the (only) crucial ingredient for quantum computation.}. It has been shown
that the principle of Bayes optimal observation \cite{Aerts Born} is
effective in both the quantum and classical observation. If only one kind of
measurement is performed, the mathematical analysis will be identical for
the quantum and classical cases. We will introduce both Bayes-optimal
detection and the popular demodulation technique and show that a comparison
of the ROC curves indicates that Bayes-optimal detection offers a vast
improvement over the latter, especially for a very low signal to noise
ratio, as in the case of NQR based TNT detection.

\section{Landmines and nuclear quadrupole resonance}

The detection of landmines turns out to be an extremely difficult task. Even
though more and more landmines are made of plastic, the bulk of landmines is
still detected using metal detectors. The reason for this, is that all
landmines contain at least a small amount of metal content in the detonator
and the metal detector gives a clear signal that is trusted by field
workers. By increasing the sensitivity of the metal detector it is possible
to reliably detect landmines. The problem is that the increased sensitivity
will make the metal detector responsive to other metal objects that abound
in postwar territory. The large false alarm rate that is accompanied by the
increased sensitivity, results on average in 500 to 1000 objects to be
wrongly classified as potential mines, for each real mine encountered. The
overhead in time, energy and cost, not to mention the high rate of
accidental detonation of real landmines as a result of this very high false
alarm rate, has spurred the search for a better classification method of the
detector signals. This classification is made more difficult by the enormous
variety of mines, soil parameters, vegetation and weather conditions. A
possible solution involves the use of nuclear quadrupole resonance
techniques. A necessary condition for the use of NQR, is the presence of a
substance with a nuclear quadrupole moment. An ideal candidate is the
naturally stable nitrogen isotope $^{14}N,$ (with a natural abundance of
99.64 \%) with nuclear spin 1 and corresponding nuclear quadrupole moment.
All known explosives contain $^{14}N,$ so that, in principle, it is possible
to detect any non-metallic mine by NQR \footnote{%
An exception is the PFM-1 landmine which contains a liquid explosive, which
is outside the scope of current NQR techniques.}. The NQR spectrum for $%
^{14}N$ has transitions in the frequency range between 0 and 6 MHz, actual
values depending mostly on the electric field gradient tensor, which is
primarily determined by the charge distribution of the electrons that bind
the nitrogen to the rest of the explosive. The resulting NQR signal is
therefore highly dependent on the chemical structure of the sample, and
delivers a potentially very reliable classification with an accompanying
very low false alarm rate. Compared with other popular mine detection
techniques such as the metal detector and the ground penetrating radar,
NQR-based detector performance is not very sensitive with respect to weather
conditions. Add to this that it is possible to construct a hand held NQR
detector, and it seems that NQR is an ideal candidate for explosive
detection \cite{Garroway}. The main challenge for NQR techniques, is the
inherently low energy content of the signal, resulting in a very low signal
to noise ratio (SNR). To improve the SNR, many repetitions of the experiment
are necessary. Rather than just measuring the free induction decay of a
single excitation, one can set up an appropriate sequence of RF pulses, and
measure the returned echo after each such pulse. In this way we obtain a
larger data set from which inferences can be made. The rate at which
repetition is physically informative, is bound from below in a fundamental
way by the physical parameters of the relaxation process. The nuclear
relaxation is a result of two different mechanisms, called the spin-spin
relaxation and the spin-lattice relaxation. The relaxation time that
characterizes the spin-lattice relaxation, denoted $T_{1}$, determines the
time necessary for the system to regain its original thermal equilibrium
state, and gives a bound on how quickly a pulse sequence can be initiated
after another. The spin-spin relaxation time, denoted $T_{2},$ is indicative
of the decoherence as a result of spin-spin interactions and determines the
length of the spin echo sequence. Spin-spin relaxation times are generally
(much) shorter than spin-lattice relaxation times. In practice, we can apply
a pulse sequence of length $T_{2},$ and repeat this pulse sequence every $%
T_{1}$. For most explosives, the relaxation times are short enough so that
NQR detection becomes feasible. Unfortunately, about 60\% of the landmines
contain $\alpha -$trinitrotoluene (TNT), which has relaxation times that
lead to prohibitively long detection times within the operational limits of
landmine detection. It is therefore projected that an NQR based landmine
detector will probably serve mainly as a confirmation sensor, i.e. a
detector that is employed to decrease the false alarm rate only after a
metal detector or a ground penetrating radar system has detected a potential
landmine. Whether used as a confirmation or as a primary detector, NQR
detection efficiency for TNT will benefit from a reduction in the time
necessary for reliable detection. Because one cannot shorten the relaxation
parameters of TNT, much effort has gone into cleverly designing the emitted
RF pulse and increasing the sensitivity of the receiver. Besides these
efforts, it is worthwhile to pursue better signal analytic detection
techniques.

\section{Quantum operations and the evolution of the NQR signal}

It is not feasible to describe the entire quantum-physical state of the
landmine, nor would this be interesting. What causes the NQR signal, is the
change in the magnetization along the direction of the solenoid. In the case
of $^{14}N$, we are dealing with a spin-1 system so that the relevant
quantum mechanical subspace is spanned by just three orthogonal vectors. A
full determination of the state in this three-dimensional subspace could, in
principle, lead to efficient (and provable optimal) strategies for detection
and classification of the NQR\ signal.\ The full characterization of the
state in just this three dimensional subspace is not feasible with a single
pulse sequence, as is the case for our data here. However, we do not
necessarily need detailed knowledge of the whole state. The mathematical
formalism required for optimal distinction between arbitrary quantum states
can easily be simplified to accommodate our limited knowledge about the
state. We will briefly show how quantum operations can serve as a framework
to relate the measured quadrature components of the current in the coil to
quantum state discrimination tools. In theoretical descriptions of NQR (\cite%
{Das and Hahn}, \cite{Lee}, and \cite{Mikhaltsevitch and Rudakov}), the
state of the system is a classical statistical mixture of pure quantum
states, described by a density operator $\rho $ belonging to the class of
linear, positive operators that sum to one when they act upon a complete set
of eigenvectors. If we consider as system the landmine, its immediate
surroundings, and the NQR detector, the detection system can be considered
as closed and the dynamics of the total density operator $\rho _{closed}$ is
governed by the unitary evolution that solves the Schr\"{o}dinger equation%
\begin{equation}
\frac{d\rho _{closed}(t)}{dt}=-\frac{i}{\hbar }[\mathcal{H},\rho
_{closed}(0)]  \label{evolution}
\end{equation}

Here $\rho _{closed}(0)$ is the initial density operator and $\mathcal{H=H}%
_{rf}\mathcal{+H}_{Q},$ with $\mathcal{H}_{Q}$ is the nuclear quadrupole
Hamiltonian, $\mathcal{H}_{rf}$ the Hamiltonian corresponding to the RF
pulse and $[$ $,$ $]$ is the commutator. The strength of the quadrupolar
Hamiltonian depends mainly on the coupling between the electric field
gradient (EFG) and the quadrupolar moment. The efficiency of the excitation
by an RF field depends on the relative orientation between the incident
radiation and the EFG principal axis frame. Because the EFG principal axis
frame depends on the molecular orientation, it is not possible to excite all
quadrupole levels with the same efficiency in a powder crystalline sample. A
calculation shows that the signal strength resulting from a crystalline
powder is approximately only 43\% the strength of a signal stemming from a
single crystal with the same number of NQR active nuclei \cite{Lee}. In
absence of the RF pulse, a canonical ensemble of NQR spin-1 systems at
temperature $T$, is described by a density operator $\rho _{thermal}:$%
\begin{equation}
\rho _{thermal}=\frac{\exp (-\mathcal{H}_{Q}/k_{B}T)}{Tr\exp (-\mathcal{H}%
_{Q}/k_{B}T)]}=\frac{1}{Z}(1-\mathcal{H}_{Q}/k_{B}T)+O(\frac{1}{T^{2}})
\label{thermal}
\end{equation}%
Here $k_{B}$ is the Boltzmann constant, $T$ the temperature in Kelvin, and $%
Z $ the partition function, which acts as a normalization. The second form
for the density operator in thermal equilibrium Eq.(\ref{thermal}), is
generally a good approximation for demining applications, as $1/k_{B}T$ is
small at room temperature in comparison to $\mathcal{H}_{Q}$. The RF pulse
perturbs the thermal equilibrium state $\rho _{thermal}$ and it is the
relaxation from this perturbed state to Eq.(\ref{thermal}), according to Eq.(%
\ref{evolution}), that is responsible for the NQR signal that we are
interested in. The Hamiltonian $\mathcal{H}_{rf}$ is $\mathcal{H}_{pulse}$
for a period of time, followed by absence of a pulse interaction for another
period of time, after which $\mathcal{H}_{pulse}$ is switched on again , and
so on. The pulse will change the EFG, and hence the coupling strength of the
quadrupole moment to the EFG. It is usual to approximate this as a series $%
\mathcal{H}_{0},$ $\mathcal{H}_{1},$ $\mathcal{H}_{2},$ $\mathcal{H}%
_{3},\ldots $ describing the Hamiltonians at the time instances $%
t_{0},t_{1},t_{2,\ldots }$ The evolution Eq.(\ref{evolution}) can then be
formally solved for $\rho $ to yield%
\begin{equation*}
\rho (t_{0}+t_{1}+\ldots )=e^{-i\mathcal{H}_{n}t_{n}}\ldots e^{-i\mathcal{H}%
_{0}t_{0}}\rho (0)e^{i\mathcal{H}_{0}t_{0}}\ldots e^{i\mathcal{H}_{n}t_{n}}
\end{equation*}%
Because our data comes from the electron current in the coil, we need a way
to connect the state of the mixture of the quadrupole active spins to this
current. The coil used is a Faraday detector, and the electron current in
the coil is the direct result of the load of the preamplifier connected to
the coil and the change of the magnetic flux inside the coil. The
expectation of the magnetization in the direction of the axis of symmetry of
the solenoid (say, the $z$-axis), is obtained by tracing over the product of
the state $\rho _{sys}$ (the mixture of quadrupole active spin-1 states)
with the magnetization operator $\mu _{z}$ along that spatial axis: 
\begin{equation}
\langle M_{z}\rangle =Tr(\mu _{z}\rho _{sys})  \label{magnetization}
\end{equation}%
Such a tracing operation, is an example of a so-called \emph{quantum
operation}. A quantum operation offers the most general possible description
of an evolution \cite{Nielsen huang}, and is defined as a mapping $%
\varepsilon $ that transforms an initial state$\rho _{0}$ to a final state $%
\rho $%
\begin{equation}
\rho =\varepsilon (\rho _{0})  \label{quantum operation}
\end{equation}%
such that there exists a set $\mathcal{O}$, called \emph{operation elements}%
, 
\begin{equation}
\mathcal{O=}\{E_{k}:\sum_{k}E_{k}E_{k}^{\dag }=I,\forall \rho :Tr(E_{k}\rho
)\geq 0\},  \label{operation elements}
\end{equation}%
for which $\varepsilon $ can be written as%
\begin{equation}
\varepsilon (\rho _{0})=\sum_{k}E_{k}\rho _{0}E_{k}^{\dag }
\label{operation representation}
\end{equation}%
The operations, by definition (\ref{operation elements}), satisfy $\sum
E_{k}E_{k}^{\dag }=I,$ and are hence \emph{trace-preserving}. Important
examples of operations that are trace-preserving, are projective
measurements, unitary evolutions and partial tracing. If the quantum
operation is a general description of a quantum measurement (or evolution),
then to each outcome $k,$ we associate one member $E_{k}$ of the collection
of measurement operators $\mathcal{O=}\{E_{k},k=1,2,...\}$ that act on the
state space. If the state is $\rho $ immediately before the measurement ,
then the probability that the outcome $k$ occurs is 
\begin{equation}
p(k|\rho )=Tr(E_{k}\rho E_{k}^{\dag })  \label{prob}
\end{equation}%
and the state after the interaction, if $k$ occurs, is 
\begin{equation}
\rho _{fin}=\frac{E_{k}\rho E_{k}^{\dag }}{Tr(E_{k}\rho E_{k}^{\dag })}
\label{final}
\end{equation}%
The two most common examples of quantum operations, are unitary
transformations ($\varepsilon (\rho _{0})=U\rho _{0}U^{\dag },$ $U$ a
unitary transformation) and von Neumann projective measurements ($%
\varepsilon _{m}(\rho _{0})=P_{m}\rho _{0}P_{m}^{\dag },$ with $P_{m}$ a
projector on the subspace labelled $m$). Many more examples, such as in
quantum computation, can be found in \cite{Nielsen huang} and modern
descriptions of quantum experiments, as in \cite{Busch et al}, \cite%
{Schroeck}. In the latter, a set $\{M_{k}\}$ of positive operators
satisfying $\sum M_{k}=I$ and $E_{k}=M_{k}^{1/2}$ is used.

Quantum operations are also a natural way to describe quantum noise and the
evolution of an open system. The mathematical prescription of a quantum
operation arises when one considers the system to be in interaction with an
environment, and that together form a closed system, for which Eq. (\ref%
{evolution}) applies. To see how this applies here, we denote the initial
state of the system under investigation by $\rho _{sys}$, and the state of
the environment (soil and interfering RF fields) as $\rho _{env},$ then the
compound system can be written as a tensor product of those states: $\rho
_{sys}\otimes \rho _{env}.$ Following the standard rules of quantum
mechanics, the expected mixture $\rho $ is the partial trace over the
degrees of the environment of the time evolved state of the closed system:%
\begin{equation}
\rho =Tr_{env}(U(\rho _{sys}\otimes \rho _{env})U^{\dag })
\label{partial trace}
\end{equation}%
It can be shown \cite{Nielsen huang} that Eq. (\ref{partial trace}) is only
slightly more general than Eq. (\ref{operation representation}), hence $\rho 
$ can be described as resulting from a quantum operation acting on the
system density matrix. Depending on whether the system contains TNT or not,
the examined system has a density matrix written as $\rho _{sys}^{tnt},$ or $%
\rho _{sys}^{1}.$ We expect either of two generic types of operation to have
occurred:%
\begin{eqnarray}
\varepsilon ^{\prime }(\rho _{sys}^{tnt}) &=&Tr_{env}(U(\rho
_{sys}^{tnt}\otimes \rho _{env})U^{\dag })=\rho ^{0}  \label{coil operations}
\\
\varepsilon ^{\prime }(\rho _{sys}^{1}) &=&Tr_{env}(U(\rho _{sys}^{1}\otimes
\rho _{env})U^{\dag })=\rho ^{1}  \notag
\end{eqnarray}%
Here $\rho ^{0}$ is the resulting mixture that produces the magnetization in
the presence of TNT, and $\rho ^{1}$ is the resulting mixture after the
interaction, in absence of TNT. An optimal detection of TNT, hence entails
optimally distinguishing the two quantum states $\rho ^{0}$ and $\rho ^{1}$.
As mentioned above, we do not posses detailed knowledge of the states $\rho
^{0}$ and $\rho ^{1}$ in practice, but we have the quadrature components $%
V(t).$ The quadrature components are a result of the change in the
magnetization $M$ Eq.(\ref{magnetization}). With $N$ the number of turns in
a solenoid of area $A$, and $Q$ the quality factor of the coil, we have%
\begin{equation}
V(t)=QN\frac{d(\mu _{0}MA)}{dt}  \label{quadrature}
\end{equation}%
We assume the quadrature components are induced by the magnetization in Eq.(%
\ref{magnetization}) by means of another quantum operation acting on the
unknown mixture $\rho _{sys}$:%
\begin{equation}
V(t)=\varepsilon _{qc}(\rho _{sys})  \label{potential operation}
\end{equation}%
Quantum operations are closed under conjunction; two consecutive quantum
operations can always be represented as a single quantum operation. What we
need to distinguish in the laboratory then, is to which type the measured $%
V(t)$ belongs:%
\begin{eqnarray}
V_{0}(t) &=&\varepsilon _{qc}(\varepsilon ^{\prime }(\rho
_{sys}^{tnt}))=\varepsilon (\rho _{sys}^{tnt})  \label{state-potential} \\
V_{1}(t) &=&\varepsilon _{qc}(\varepsilon ^{\prime }(\rho
_{sys}^{1}))=\varepsilon (\rho _{sys}^{1})  \notag
\end{eqnarray}%
Quantum operations can only have the effect of reducing the trace distance,
which in turn will increase the minimal Bayes risk associated with
distinguishing the two situations. Hence Bayes optimal detection using the
quadrature components induces some loss in detector performance in
comparison with the same procedure applied to a reconstruction of the state $%
\rho _{sys}$; because we skip one quantum operation in Eq.(\ref%
{state-potential}), this would lead to a lower Bayes-risk.

\section{Detection schemes}

\subsection{Bayes optimal observation of NQR data}

In essence, Bayes-optimal detection deals with the optimal decision of a
hypothesis from a set of mutually exclusive hypotheses. Consider the binary
decision problem 
\begin{eqnarray*}
H_{0} &:&\text{the signal indicates TNT presence} \\
H_{1} &:&\text{the signal indicates no TNT presence}
\end{eqnarray*}%
If a given set of data is compatible only with one of the two hypotheses,
the decision problem becomes trivial. However, in practice, the data
generally supports both hypotheses, albeit with a different probability, and
the decision task is consequently complicated by this fact. If we are given
data $x_{i}$ from a set of possible outcome results $X=\{x_{1},x_{2},\ldots
,x_{i},\ldots ,x_{n}\},$ and the factual occurrence of $x_{i}$ supports both
hypotheses, we need to infer what the probability was of getting the result $%
x_{i}$ as a result of either hypothesis being true. That is, we need some
means to evaluate $p(x_{i}|H_{0})$ and $p(x_{i}|H_{1}).$ Any additional
(prior) information can be included under the label $D$ and then we compare $%
p(x_{i}|H_{0},D)$ and $p(x_{i}|H_{1},D).$ Of course, what we are after, is
the probability of $H_{0}$ or $H_{1}$ being true, on the condition that $D$
holds and $x_{i}$ was the outcome of the experiment. By the use of Bayes'
theorem \cite{Jaynes2003}, we have 
\begin{equation}
p(H_{0}|x_{i},D)=p(H_{0}|D)\frac{p(x_{i}|H_{0},D)}{p(x_{i}|D)}
\label{Bayes H0}
\end{equation}%
\begin{equation}
p(H_{1}|x_{i},D)=p(H_{1}|D)\frac{p(x_{i}|H_{1},D)}{p(x_{i}|D)}
\label{Bayes H1}
\end{equation}%
We eliminate the denominator by calculating the ratio of Eq. (\ref{Bayes H0}%
) and Eq. (\ref{Bayes H1}):%
\begin{equation}
\frac{p(H_{0}|x_{i},D)}{p(H_{1}|x_{i},D)}=\frac{p(H_{0}|D)}{p(H_{1}|D)}\frac{%
p(x_{i}|H_{0},D)}{p(x_{i}|H_{1},D)}  \label{ratio}
\end{equation}%
In absence of any preference which of the two hypotheses is more likely than
the other on the basis of the prior information, we set $\frac{p(H_{0}|D)}{%
p(H_{1}|D)}=1$. In complete absence of any prior information, we omit
dependence on $D.$ The quantity of interest for optimally choosing between
two alternative hypotheses, is the likelihood ratio (also called \emph{the
odds} in the binary case):%
\begin{equation}
\Lambda _{i}=\frac{p(H_{0}|x_{i})}{p(H_{1}|x_{i})}  \label{odds}
\end{equation}

We call the observation scheme \emph{Bayes-optimal} iff the obtained outcome 
$x_{i}$ is the outcome that maximizes the odds, Eq.(\ref{odds}), that the
outcome given, pertains to the system under investigation rather than to
noise in the detection system. It turns out that this is a model for quantum
as well as classical observation \cite{Aerts S}. Assuming our detector is
Bayes-optimal, allows for an optimal detection strategy by reversing the
logic of the detector\footnote{%
A very similar approach to observation with the same name, is proposed in
several papers that deal with visual perception by humans. We refer to \cite%
{Geisler Kersten} and the references found there.}. Of course, we do not
know in advance whether the physical detector satisfies the condition of
Bayes-optimality, and actual performance will depend on how well this
condition will be met. In accordance with quantum mechanics, we assume the
probability $p(H_{0}|x_{i})$ (and $p(H_{1}|x_{i})$) is a monotone function
of the trace distance between the actually measured signal, and the ideal
(averaged over many samples) signal obtained in the presence (absence) of
TNT. Numerator and denominator in Eq.(\ref{odds}) can be substituted by the
corresponding trace distance, as the outcome for which the likelihood ratio
is maximal, is invariant under monotone transformations. A second rationale
for taking the trace distance, is that it arises naturally when one
considers the Bayes risk in the binary state discrimination problem.

\subsection{Trace distance and Bayes risk of distinguishing quantum states}

If we are given two states $\rho ^{0}$ and $\rho ^{1}$ with \emph{a priori}
probabilities $p_{0}$ and $p_{1}=1-p_{0},$ then, following Eq. (\ref{quantum
operation}), we look for two operations elements $\mathcal{O=}%
\{E_{0},E_{1}\} $ such that $E_{0}+E_{1}=I$ and $E_{0},E_{1}\geq 0$ that
minimize the \emph{Bayes risk} or probability of error \cite{Helstrom} :%
\begin{equation}
R_{\mathcal{O}}(p_{0})=p_{0}Tr(\rho ^{0}E_{1})+p_{1}Tr(\rho ^{1}E_{0})
\label{Bayes risk}
\end{equation}%
rewriting Eq.(\ref{Bayes risk}) once with $E_{1}=I-E_{2}$ and once with $%
E_{2}=I-E_{1},$ adding and dividing, yields%
\begin{equation*}
R_{\mathcal{O}}(p_{0})=\frac{1}{2}[1-Tr[(p_{0}\rho ^{0}-p_{1}\rho
^{1})(E_{0}-E_{1})]
\end{equation*}%
To proceed, we define the trace distance between $\rho ^{0}$and $\rho ^{1}$,
as 
\begin{equation}
D(\rho ^{0},\rho ^{1})=\frac{1}{2}Tr\sqrt{(\rho ^{0}-\rho ^{1})(\rho
^{0}-\rho ^{1})^{\dag }}  \label{trace distance}
\end{equation}%
An important property of the trace distance is that it is symmetric in its
arguments, positive iff $\rho ^{0}\neq \rho ^{1},$ zero iff $\rho ^{0}=\rho
^{1},$ and satisfies the triangle inequality. In other words, it is a
bona-fide distance measure on the set of density matrices. Another important
property of the trace distance, is given by 
\begin{equation*}
D(\rho ^{0},\rho ^{1})=\max_{E_{i}\in \mathcal{O}}Tr(E_{i}(\rho ^{0}-\rho
^{1}))
\end{equation*}%
With this we can show \cite{Nielsen huang} that the minimum value the Bayes
risk $\min_{\mathcal{O}}R_{\mathcal{O}}(p_{0})$ can attain does not depend
on the $E_{k}$ and equals 
\begin{gather}
\min_{\mathcal{O}}R_{\mathcal{O}}(p_{0})=R_{Bayes}(p_{0})
\label{minimal risk} \\
=\frac{1}{2}-\frac{1}{2}Tr[\sqrt{(p_{0}\rho ^{0}-p_{1}\rho ^{1})(p_{0}\rho
^{0}-p_{1}\rho ^{1})^{\dag }}
\end{gather}%
In our treatment of the data, each specific sample could equally well
contain TNT, or not, so that we have as prior probabilities $p_{0}$ $%
=p_{1}=1/2:$ 
\begin{equation}
R_{Bayes}(p_{0})=\frac{1}{2}-D(\rho ^{0},\rho ^{1})  \label{bayes-risk}
\end{equation}%
We see the minimal Bayes risk is attained for two states that maximize the
trace distance. Trace preserving quantum operations can be shown to cause a
contraction in the space of density operators \cite{Nielsen huang}. Because
the trace distance is a true distance measure on the space of density
operators, it can only decrease as a result of an arbitrary trace-preserving
quantum operation $\varepsilon $:%
\begin{equation}
D(\rho ^{0},\rho ^{1})\geq D(\varepsilon (\rho ^{0}),\varepsilon (\rho ^{1}))
\label{contractivity}
\end{equation}%
If the current in the coil is the result of Eq. (\ref{coil operations}),
then being able to distinguish the currents reliably (i.e., the trace
distance is greater than can be explained from fluctuations), indicates we
have successfully distinguished the situations represented by $H_{0}$ and $%
H_{1}.$Quantum operations can only have the effect of reducing the trace
distance, which in turn will increase the minimal Bayes risk associated with
distinguishing the two situations. Hence Bayes optimal detection using the
quadrature components induces some loss in detector performance in
comparison with the same procedure applied to a reconstruction of the state $%
\rho _{sys}$; because we skip one quantum operation in Eq.(\ref%
{state-potential}), this would lead to a lower Bayes-risk.

\subsection{The Demodulation technique}

A popular method to establish the presence of a given substance in a NQR
tested sample, is the use of the so-called demodulation technique. This
method is particularly simple and consists of calculating an estimate $%
\sigma (\nu _{n})$ of the power spectral density $S(\nu )$ of the signal $%
s(t_{n}),$ $n=1,\ldots ,256,$ by first fast Fourier transforming the signal
and taking its modulus squared. Let us call $\nu _{\max }$ the frequency $%
\nu _{\max }=\arg (\max (S(\nu )))$ where one expects the spectral line with
the highest intensity in presence of TNT. The value of the \emph{estimated}
power spectral density $\sigma (\nu _{\max })$ evaluated at the frequency $%
\nu _{\max },$ is then the test statistic for a treshold detector. If $%
\sigma (\nu _{\max })$ exceeds a given treshold, the presence of TNT is
accepted, if not, it is rejected. The estimated $\sigma (\nu _{n})$ will in
general deviate from $S(\nu )$ at the precise values $\nu _{n}$, but may be
approximately regarded as an average over the interval $[\frac{\nu _{n}-\nu
_{n-1}}{2},\frac{\nu _{n+1}-\nu _{n}}{2}].$ To account for this, sometimes
the average under $\sigma (\nu _{n})$ over a few frequency bins is taken as
a statistical test parameter. Whether this is useful depends, among other
things, on the magnitude of the width of the spectral line with the highest
intensity relative to the width of the frequency bins. Moreover, as NQR
spectra are generally a function of the temperature of the sample, and
because this parameter is difficult to estimate in demining applications
within a range of 5 to 10 Kelvin, the value of $\nu _{\max }$ will depend on
the temperature too. To make sure we do not miss the peak, one can then take
the area over a region in the frequency domain where one expects the peak.
This complication presents no real problem to the method employed and,
however important to the actual demining problem, is not taken into account
here (see, however, \cite{Andreas et al} and \cite{Mikhaltsevitch and
Rudakov}). All our experimental samples are taken at the same temperature.
As expected, we see little change in the efficiency of the method, whether
we use $\sigma (\nu _{\max }),$ or a sum of values $\sum \sigma (\nu _{\max
})$ for a tiny region surrounding the relevant frequency bin. However, a
considerable improvement is obtained when we allow for the demodulation
technique to sample multiple peaks simultaneously. The results that we
present here, employ the single peak value $\sigma (\nu _{\max })$ of the
frequency bin containing the mean excitation frequency 841.5 kHz, as well as
an improved demodulation algorithm exploiting knowledge of three resonance
frequencies of TNT within a range of a few tens of kHz around the mean
excitation frequency 841.5 kHz.

\section{ Experimental results}

\subsection{Set up and data acquisition}

The data employed for our analysis was kindly provided by the NQR group of
King's College, London, under supervision of Professor J.A.S. Smith. In the
experimental set up employed, a pure monoclinic TNT\ sample with a weight
typical of that found in an anti-personnel mine, is placed inside a
solenoidal coil. The same coil is used for emission of the RF-pulse, as well
as for the reception of the subsequent echo. The returned echo signal is
routed through a hardware band-pass filter with a bandwidth of approximately
50 kHz and subsequently sent to a Tecmag Libra spectrometer that splits the
signal in two signals which are then mixed with two quadrature components,
yielding a complex discrete time series. Because of the ideal laboratory
conditions under which the signal has been obtained, the results will
compare unrealistically optimistic with respect to those obtained under
field conditions. In particular, the absence of RF interference and the use
of a coil that contains the sample in its entirety, must be taken into
account when attempting to compare the results of our analysis with those of
data obtained under more realistic conditions. The emitted RF signals are
pulsed, spin locked echo signals with a mean excitation frequency of 841.5
kHz.\ The mean and width of the excitation are such, that 4 spectral lines
of TNT\ can be detected within the frequency range of the band pass filter.
Because the same coil is used for the emission of the RF signal (which has a
mean power of several kilo watts), as for the reception of the echo (which
is extremely weak), the returned echo contains so-called antenna ringing
effects. To cancel the effect of the antenna ringing, a phase cycling
technique, popular in the more established field of NMR, is employed. The
phase cycling technique requires forming an appropriate sum of four signals.
The signals used for the analysis, are the sum of 5 such phase-cycled sums
and hence consist of 20 repeated data acquisitions, averaged to improve
signal to noise ratio. The sampling time is 5%
$\mu$%
S and each set has 8192 data points, which consists of 32 sequential echo
signals, each containing 256 data points. The pulse sequence is of the type 
\begin{equation*}
\pi -\tau -\pi -2\tau -\pi -2\tau -\pi -2\tau -...
\end{equation*}

Here $\pi $ denotes the RF pulses and the 1280 
$\mu$%
S of data (256 times 5 
$\mu$%
S) for each echo signal is acquired during the $2\tau $ periods between the
pulses. All algorithms are programmed in MATLAB 7 on a 2,2 GHz PC with 512
MB RAM. Both methods are fast: the determination of whether a given signal
was obtained in the presence of TNT or not, requires in both cases a
calculation time less than a second, more than one order of magnitude below
the necessary data acquisition time.

\subsection{Detector performance}

The statistical assessment of detector performance is based on the
sensitivity, specificity and ultimately on the functional relationship which
exists between these two, as expressed in the receiver operating
characteristic (ROC). The true positive rate, or sensitivity, is the
probability that the detector indicates the presence of a mine, when there
is indeed a mine present. The false positive rate is defined as the
probability that the detector indicates the presence of a mine when there
was no mine present. The specificity is then defined as $1-$false positive
rate. Increasing the sensitivity of the detector, lowers the specificity and
vice versa. The overall performance is therefore conveniently expressed by
means of a receiver operating characteristic (ROC) curve, which plots the
sensitivity as a function of the false positive rate. The data used to
calculate the ROC curves, consist of 100 data samples with TNT, and 100 data
samples without TNT. Because of spin-lattice relaxation, we expect the
signal quality to decrease as a function of the echo number, a behavior we
see reflected in the ROC curves and in the line intensities of the three
most visible resonances as a function of the echo number, as depicted in
figure \ref{pict0}. \FRAME{ftbpFU}{3.9991in}{2.4369in}{0pt}{\Qcb{%
{\protect\small The intensity of the three TNT quadrupole resonances within
experimental reach, for a sum of 100 signals, as a function of the
echonumber. The decrease in intensity is in good approximation exponential.
The intensity of the last few echoes is two orders of magnitude smaller than
the first.}}}{\Qlb{pict0}}{intensity_of_resonances2.eps}{\special{language
"Scientific Word";type "GRAPHIC";maintain-aspect-ratio TRUE;display
"USEDEF";valid_file "F";width 3.9991in;height 2.4369in;depth
0pt;original-width 15.8613in;original-height 9.6193in;cropleft "0";croptop
"1";cropright "1";cropbottom "0";filename 'E:/Matlab/work/Pictures BO
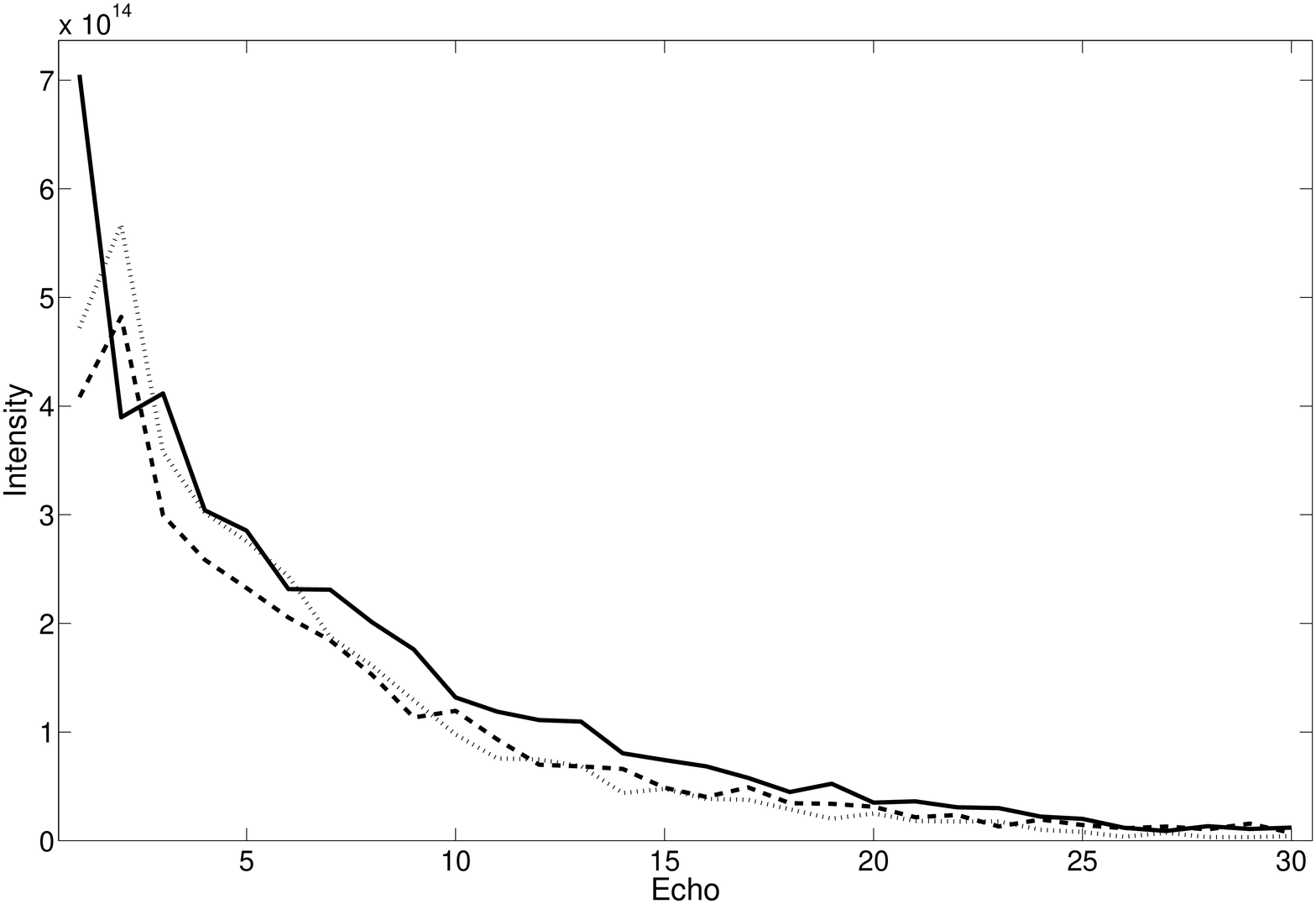';file-properties "XNPEU";}}

In figure \ref{pict1} and figure \ref{pict2}, we have depicted ROC curves
for both detection methods. As is well-known, ROC curves can be
\textquotedblleft convexified\textquotedblright\ using mixed measurements,
i.e., measurements that are linear combinations of measurements with a
threshold value that corresponds to extremal points of the experimental ROC
curve. This procedure results in an improved detector. Nevertheless, we have
depicted all ROC curves calculated only for single threshold values; the
convexified ROC curves can easily be visualized from the given curves. 
\FRAME{ftbpFU}{3.9914in}{2.4326in}{0pt}{\Qcb{{\protect\small ROC curves for
the demodulation technique for a single peak. In decreasing order of
performance (ever lower ROC curves), we plotted ROC curves for echoes 9
(solid), 13 (dashes), 17 (point-dash) and 21 (points) respectively. The
first four echoes yield close to perfect detectors.}}}{\Qlb{pict1}}{%
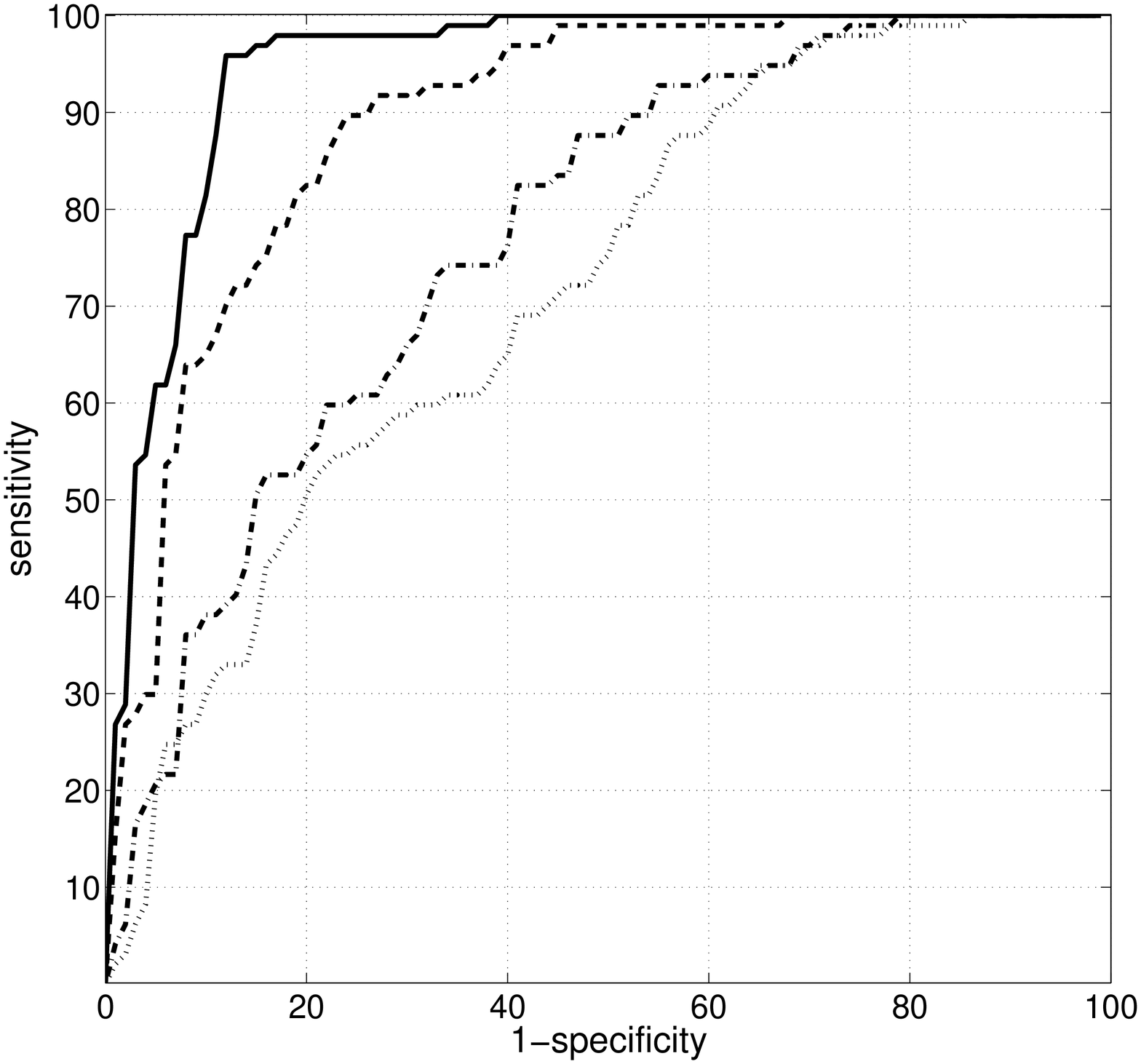}{\special{language "Scientific Word";type
"GRAPHIC";maintain-aspect-ratio TRUE;display "USEDEF";valid_file "F";width
3.9914in;height 2.4326in;depth 0pt;original-width 15.8613in;original-height
9.6193in;cropleft "0";croptop "1";cropright "1";cropbottom "0";filename
'E:/Matlab/work/Pictures BO
TNT/demodulation/onepeak_new2.eps';file-properties "XNPEU";}} \FRAME{ftbpFU}{%
3.9914in}{2.4326in}{0pt}{\Qcb{{\protect\small ROC curves for the Bayes
optimal detector. As in the previous graph, shown are ROC curves for echo
numbers 9 (solid), 13 (dashes), 17 (point-dash) and 21 (points), in
decreasing order of performance respectively. The first ten echoes yield
close to ideal detectors. }}}{\Qlb{pict2}}{bayesfourechoes6.eps}{\special%
{language "Scientific Word";type "GRAPHIC";maintain-aspect-ratio
TRUE;display "USEDEF";valid_file "F";width 3.9914in;height 2.4326in;depth
0pt;original-width 15.8613in;original-height 9.6193in;cropleft "0";croptop
"1";cropright "1";cropbottom "0";filename 'E:/Matlab/work/Pictures BO
TNT/ROC 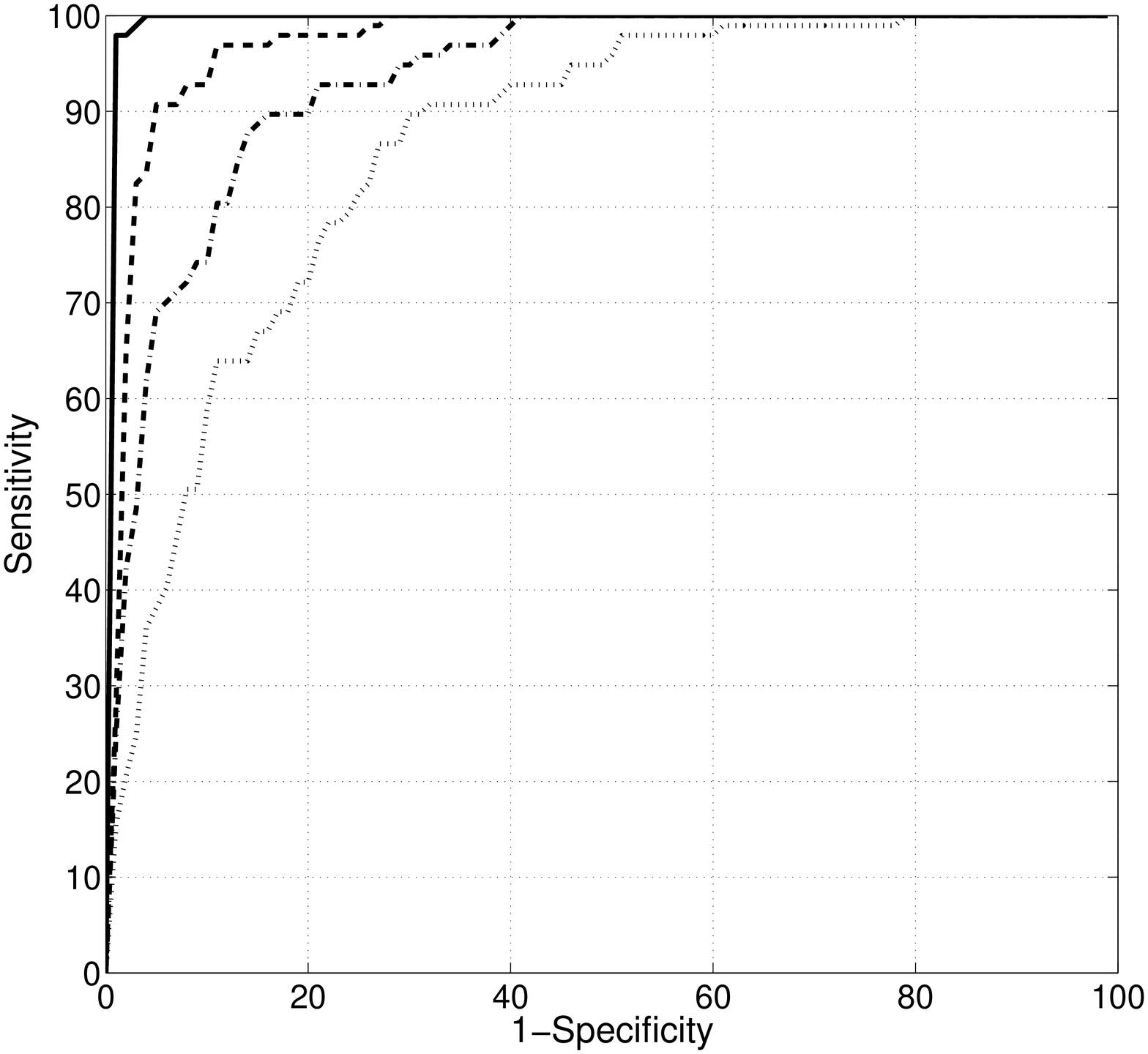';file-properties "XNPEU";}}The Bayes
optimal detector uses the whole signal without noise reduction (except for
the hardware band-pass filter), whereas the demodulation detector measures
the intensity of the single peak with the highest intensity, which is the
same as filtering the signal with a very narrow boxcar band-pass filter. In
this sense the two methods are opposites of one another. An intermediate
detector can be obtained by using the three most distinct resonances in the
NQR spectrum of TNT within the frequency band allowed by the band-pass
filter. As can be seen from the ROC curves in figure \ref{pict3}, the
performance of the three peak demodulation technique, although still lagging
behind the Bayes optimal detector, is better than the single peak detector. 
\FRAME{ftbpFU}{3.9914in}{2.4369in}{0pt}{\Qcb{{\protect\small An improved
demodulation detector utilising the three most distinct peaks in the NQR\
spectrum. As before, depicted are echoes 9 (solid), 13 (dashes), 17
(point-dash) and 21 (points). One can see the detector is a considerable
improvement over the single peak detector, but still lags behind the Bayes
optimal detector.}}}{\Qlb{pict3}}{roc_4echoes_3peaks.eps}{\special{language
"Scientific Word";type "GRAPHIC";display "USEDEF";valid_file "F";width
3.9914in;height 2.4369in;depth 0pt;original-width 15.8613in;original-height
9.6193in;cropleft "0";croptop "1";cropright "1";cropbottom "0";filename
'E:/Matlab/work/Pictures BO TNT/improved
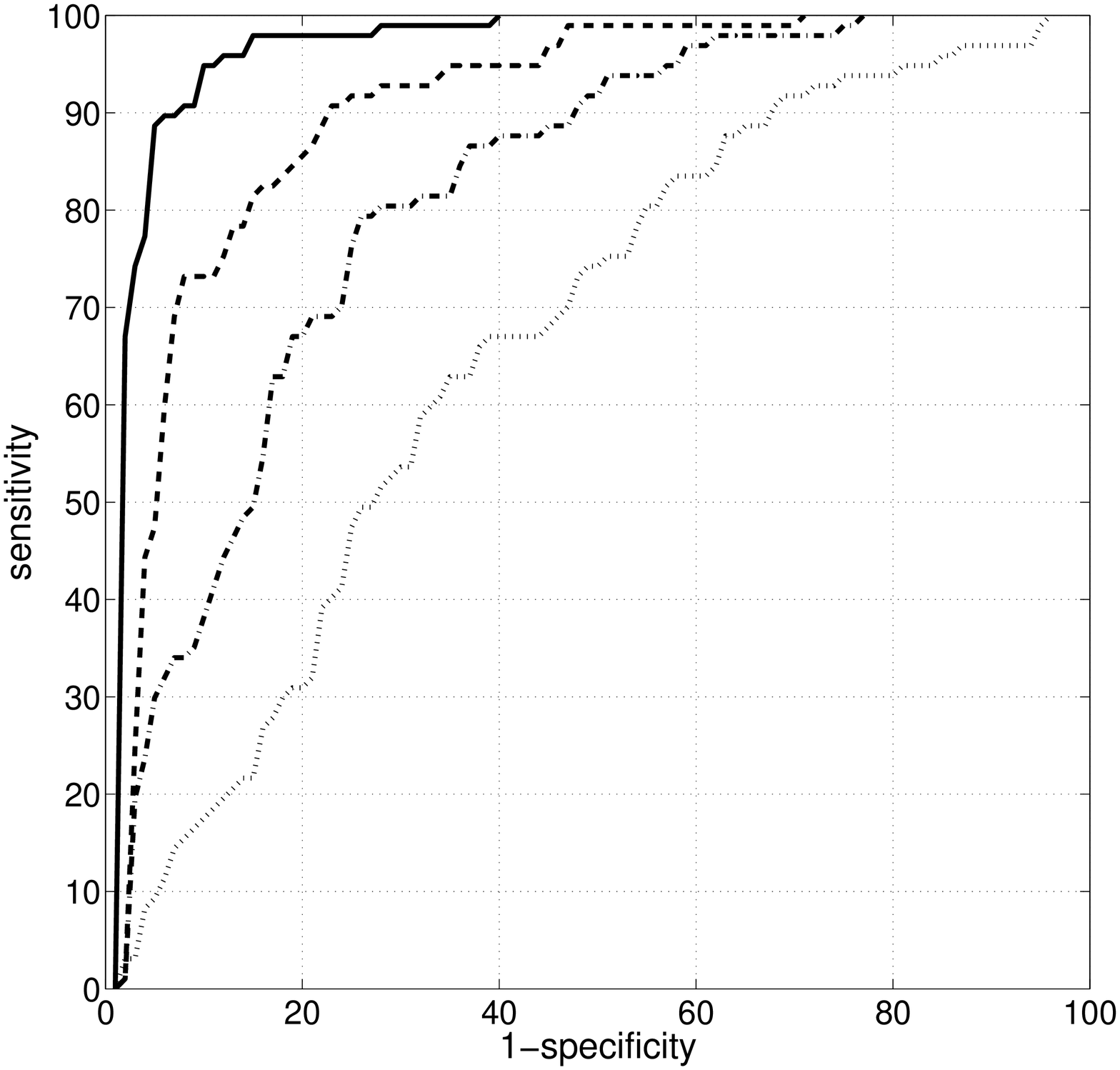';file-properties "XNPEU";}}The
normalized area under a ROC curve can be taken as a crude measure of the
overall performance of the detection scheme. The ideal case corresponds to
an area of one, the completely ignorant detector scores one half. Due to the
high risks involved in the practice of demining, it is crucial no mine be
missed, i.e., one wants to have a sensitivity as close to 100\% as possible. 
\FRAME{ftbpFU}{3.5763in}{2.155in}{0pt}{\Qcb{{\protect\small A comparison
between the overall performance of the demodulation (dotted curve) and Bayes
optimal detection (solid curve) methods. Depicted is the area under the ROC
curve as a function of the echo number. One can see how for both methods the
performance decreases with increasing echo number. The Bayes optimal
detection clearly outperforms the demodulation technique after echo number
4. }}}{\Qlb{pict4}}{arearoc3.eps}{\special{language "Scientific Word";type
"GRAPHIC";display "USEDEF";valid_file "F";width 3.5763in;height
2.155in;depth 0pt;original-width 15.8613in;original-height 9.3443in;cropleft
"0";croptop "1";cropright "1";cropbottom "0";filename
'E:/Matlab/work/Pictures BO TNT/Area of ROC for all single
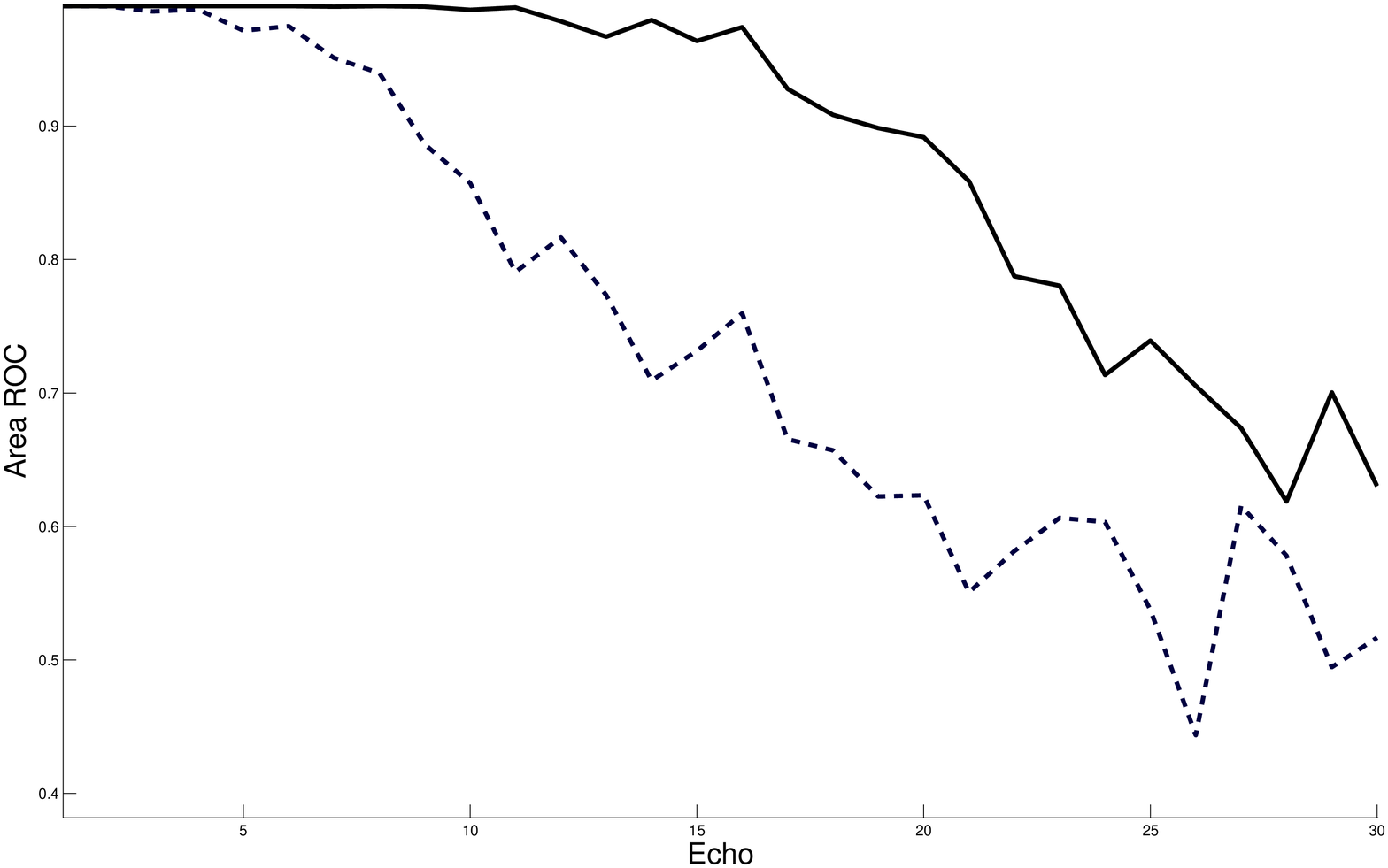';file-properties "XNPEU";}}The Bayes optimal detector
allows to use any of the first 10 echo numbers to obtain a detector that is
very close to ideal. In contrast, the demodulation detector only yields a
close to ideal detector for the first 4 echoes. We see in figure \ref{pict4}
that the overall performance of the demodulation detector decreases much
more rapidly than the Bayes optimal detector. As the first echo already
yields a perfect detector for both methods, there seems no obvious incentive
to improve the detection capabilities. However, the spin-lattice relaxation
constrains the time between the spin-locked pulses to a minimum of 10
seconds. Hence the necessary data acquisition time for the each single data
sample, is approximately 20 acquisitions*10 seconds = 200 seconds. In actual
demining applications, the necessary acquisition time will further increase
as a result of RF interference, other NQR active soil constituents such as
piezoelectric ceramics, and the fact that only single sided (as opposed to
the sample being \emph{within }the coil, as is the case for our data),
remote acquisition is possible. However, one can substantially decrease this
acquisition time by combining the information in the different echoes. It is
hence of vital importance to improve the detector performance for \emph{all}
the echoes in the pulse sequence. The proposed detector succeeds in doing
just that.

\section{Concluding remarks}

We examined the applicability of a simple Bayes-optimal quantum state
discrimination technique to see if it is possible to improve the detection
capabilities of remote TNT\ detection by NQR measurements. Although the
experimental setup employed here is only able to give an estimate of a
projection of the spin state onto the axis of symmetry of the solenoid, the
method delivers a very reliable detector. A comparison was made with the
popular demodulation detector. Both methods are simple and fast, but our
results indicate the Bayes-optimal scheme offers a considerable improvement
over the demodulation approach. An extension of the demodulation approach,
in which the three most intense peaks are combined, offers an improvement
with respect to the single frequency demodulation, but is still considerably
less performant than the Bayes optimal scheme. The difference in performance
between the two methods becomes greater as the echo number increases. For
the last few echoes, the advantage becomes less pronounced, which we
attribute to the fact that the signal strength of the echo diminishes
exponentially with increasing echo number, so that eventually both methods
will fail to deliver for very weak echoes. Handling signals with a low SNR
is important, as one expects a deterioration of the already low SNR inherent
in NQR measurements in actual field tests. The proposed detector offers two
distinct and important advantages with respect to demodulation for demining
applications: increase of the specificity (without sacrificing the close to
perfect sensitivity necessary for demining applications), and a decrease of
the necessary detection time. It is possible to include data from primary
detectors (such as a metal detector or ground penetrating radar) in the form
of prior probabilities, so that the NQR detector becomes a confirmation
sensor. It would be interesting to combine different pulse sequences that
allow for a more complete reconstruction of the full density matrix of the
spin-1 NQR system, and see whether this leads to a better detector as a
result of the further decreased minimal Bayes risk. Varying the pulse scheme
may offer other advantages too. By tailoring the pulse sequences to enact on
disjoint excitations in the frequency plane, one may, depending on the
magnitude of the cross-relaxation between the modes, be able to improve the
extraction of the information content by questioning different modes. If
different pulse sequences are transmitted by different antennae, one can
improve the fraction of nuclei participating in the NQR signal above the
43\% limit for a single field orientation. As the state contains all
information about the system, a detector based on the reconstructed density
operator, yields an approximation to a truly optimal detector. It remains to
be seen whether an implementation of such a detector offers practical
improvements for demining applications.

\section{Acknowledgements}

This work was done as part of Flemish Fund for Scientific Research (FWO)
research project G.0362.03N. We gratefully acknowledge J.A.S. Smith and M.
Rowe for supplying us their NQR data and helpful feedback on our results. We
thank Hichem Sahli, Luc van Kempen, Andreas Jakobsson and Sam Somasundaram
for discussions and feedback.

\end{document}